\documentclass[aps,pra,twocolumn,superscriptaddress,amsmath,amsfonts,amssymb,preprintnumbers,bibnotes]{revtex4-1}
\usepackage[T1]{fontenc}
\usepackage{wasysym}
\usepackage[latin1]{inputenc}
\usepackage{graphicx}
\usepackage[colorlinks,plainpages=false,linkcolor=black,urlcolor=blue,citecolor=black,pdfpagemode=UseNone,pdfstartview=FitBH]{hyperref}

\makeatletter
\renewcommand{\@evenfoot}{\hfill\raisebox{-3em}{\bf\thepage}\hfill}
\renewcommand{\@oddfoot}{\hfill\raisebox{-3em}{\bf\thepage}\hfill}
\makeatother

\begin{document}

\title{Spin phases of the helimagnetic insulator Cu$_2$OSeO$_3$ probed by magnon heat conduction}
\author{N.~Prasai}
\affiliation{Department of Physics, University of Miami, Coral Gables, FL 33124}
\author{A.~Akopyan}
\affiliation{Department of Physics, University of Miami, Coral Gables, FL 33124}
\author{B.~A. Trump}
\altaffiliation[Current address: ]{NIST Center for Neutron Research, National Institute of Standards and Technology, Gaithersburg,
Maryland 20899-6102}
\affiliation{Department of Chemistry, Johns Hopkins University, Baltimore, MD 21218}
\affiliation{Department of Physics and Astronomy, Institute for Quantum Matter, Johns Hopkins University, Baltimore, MD 21218}
\author{G.~G. Marcus}
\affiliation{Department of Physics and Astronomy, Institute for Quantum Matter, Johns Hopkins University, Baltimore, MD 21218}
\author{S.~X. Huang}
\affiliation{Department of Physics, University of Miami, Coral Gables, FL 33124}
\author{T.~M. McQueen}
\affiliation{Department of Chemistry, Johns Hopkins University, Baltimore, MD 21218}
\affiliation{Department of Physics and Astronomy, Institute for Quantum Matter, Johns Hopkins University, Baltimore, MD 21218}
\affiliation{Department of Material Science and Engineering, Johns Hopkins University, Baltimore, MD 21218}
\author{J.~L.~Cohn}
\email[Corresponding Author: ]{cohn@physics.miami.edu}
\affiliation{Department of Physics, University of Miami, Coral Gables, FL 33124}

\begin{abstract}
We report studies of thermal conductivity as functions of magnetic field and temperature in the helimagnetic insulator
Cu$_2$OSeO$_3$ that reveal novel features of the spin-phase transitions as probed by magnon heat conduction. The tilted
conical spiral and low-temperature skyrmion phases, recently identified in small-angle neutron scattering studies, are
clearly identified by sharp signatures in the magnon thermal conductivity. Magnon scattering associated with the presence
of domain boundaries in the tilted conical phase and regions of skyrmion and conical-phase coexistence are identified.
\end{abstract}

\maketitle
\clearpage


The cubic chiral magnets (MnSi, FeGe, Cu$_2$OSeO$_3$) have attracted considerable attention for their complex variety of non-collinear
spin phases that include spin modulations with long periods (many lattice spacings) and topological skyrmion phases. Their similar and
rich magnetic phase diagrams are dictated by their common noncentrosymmetric cubic lattice symmetry and
a hierarchy of competing energy scales (e.g. exchange, Dzyaloshinsky-Moriya, magnetocrystalline anisotropy, Zeeman). Recently two
new spin phases, low-temperature skyrmion and  ``titled conical spiral,'' were identified in the insulating compound Cu$_2$OSeO$_3$
by small-angle neutron scattering (SANS) \cite{Chacon,Qian}. These novel phases, arising
at low temperature and relatively high magnetic field, reflect competing Zeeman and anisotropy energies that lead to surprising
spin textures and a re-orientation of the long-period spin modulation direction.

Here we report that field dependent thermal conductivity measurements are a particularly sensitive probe of the spin phase transitions in
Cu$_2$OSeO$_3$ because of the compound's unprecedentedly large magnon heat conductivity \cite{PrasaiK}.
Rather little is known experimentally about magnons in ferro- or ferri-magnets from heat transport, their scattering or
its dependency on spin textures, whether long-range ordered or not. Such information has increased in its importance and
relevance with the surging interest in spintronic and magnonic device applications \cite{SpinCaloritronics,MagnonSpintronics}.
Cu$_2$OSeO$_3$ is an ideal material for investigating these characteristics because the spin and lattice systems are weakly coupled as
evidenced by very low spin-lattice damping \cite{Seki,NewAPL} and by mean-free paths for both magnons and phonons as large as 0.3~mm
below 2K \cite{PrasaiK}. These conditions ensure that energy (e.g. from a heater in thermal conductivity measurements) is transferred
from the lattice to the spin system, but is weak enough so the contributions from phonons ($\kappa_L$) and magnons ($\kappa_m$) are
approximately additive \cite{SandersWalton,PrasaiK}, $\kappa\simeq \kappa_L+\kappa_m$.

Our measurements reveal a complete suppression of the magnon heat flux in the tilted conical phase along the $\langle 110\rangle$ directions
that we attribute to strong magnon scattering by tilt domain boundaries. This observation raises the prospect of exploiting this configuration
of heat flux and applied field in a field-controllable spin-current switch. The low-temperature skyrmion phase, characterized by long-range
skyrmion lattice order, supports maximum magnon heat conduction comparable to that of the fully polarized phase. Suppressed magnon heat
conduction characterizes regions of phase coexistence.

Cu$_2$OSeO$_3$ comprises a three-dimensional distorted pyrochlore (approximately fcc) lattice of corner-sharing Cu tetrahedra \cite{CrystalSymmetry,CrystalSymmetry2}. Strong magnetic interactions within tetrahedra lead to a 3-up-1-down, spin $S=1$ magnetic state \cite{Palstra,Belesi} with weaker interactions between tetrahedra leading to their ferromagnetic ordering below $T_C\simeq 58$~K.
At low temperatures \cite{Skyrmion,SpecHeat} the low-field state is helimagnetic wherein the atomic spins rotate within a plane perpendicular to the helical axis
with a wavelength $\lambda_h\simeq 62$~nm; multiple domains with helices aligned along $\langle 100\rangle$ directions (easy axes) characterize this phase. At $H\gtrsim 10-25$~mT (depending on field orientation) the helices of individual domains rotate along the field to form a single-domain, conical phase in which spins rotate on the surface of a cone. Further increasing the field narrows the conical angle until $H\gtrsim 50-75$~mT where the ferrimagnetic, fully-polarized (collinear) state emerges. Until recently, these three phases comprised a universal phase diagram for the cubic chiral magnets.
\begin{figure*}[t]
\includegraphics[width=5.2in,clip]{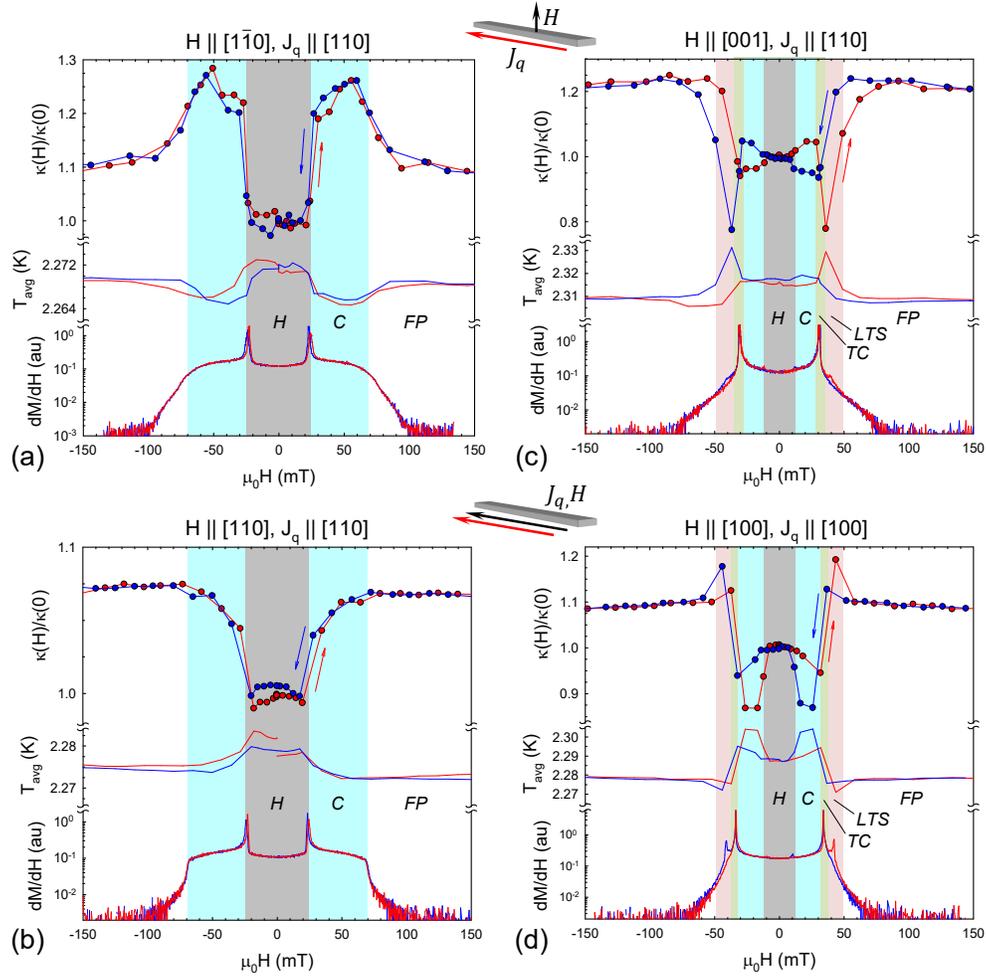}
\caption{
(a)-(d): $\kappa$, $T_{avg}$ and $dM/dH$ vs field (corrected for demagnetization) at $T\simeq 2.30$~K for four configurations of the applied
field and heat flow. $H\perp J_q$ for (a) and (c), $H|| J_q$ for (b) and (d). Specimens were cooled in zero field and hysteresis loops traversed
sequentially from (arrows, upper panel): $\mu_0H=0\to 200$~mT, $200$~mT~$\to -200$~mT, $-200$~mT~$\to 0$.
The spin phases are distinguished by different background shading and labels: helical (\emph{H}), conical ({\emph C}), tilted conical (\emph{TC}),
low-temperature skyrmion (\emph{LTS}), and fully polarized (\emph{FP}).}
\label{Fig1}
\end{figure*}

Our prior studies of this compound for heat flow along $[111]$ [\emph{cf.} Fig.~\ref{Fig3}~(e)]
demonstrate that the field dependent changes in $\kappa$ are entirely magnonic. A separation of $\kappa_L$ and
$\kappa_m$ is possible at high fields since spin-wave excitations are gradually depopulated (gapped) with increasing field following the Zeeman
energy $E_H={\textsl g}\mu_B H$, and are completely suppressed for $E_H\gg k_BT$ where $\kappa\to
\kappa_L$ \cite{FriedbergDouthett,McCollum,Luthi,Douglass,BhandariVerma,WaltonRives,Pan,BoonaHeremans}.
This decomposition, at $T\lesssim 1.2$~K where the high-field condition applies, is discussed in detail for [111]-oriented
specimens in Ref.~\onlinecite{PrasaiK}. That work demonstrates mean-free paths for both phonons and magnons comparable to
the transverse specimen dimensions at $T\lesssim 2$~K and maximum values for $\kappa_m\sim 20-80$~W/mK, the largest known
for ferro- or ferri- magnets. Values for $\kappa_L$ and $\kappa_m$ vary modestly between specimens due to size effects and
a small and variable Se vacancy concentration \cite{PrasaiK}.

Phase pure, single crystals of Cu$_2$OSeO$_3$ were grown by chemical vapor transport as described previously \cite{Growth}. Specimens were cut from single-crystal ingots, oriented by x-ray diffraction, and polished into thin parallelopipeds. A two-thermometer, one-heater method was employed to measure the thermal conductivity in applied magnetic
fields up to 50 kOe. Specimens were suspended from a Cu heat sink with silver epoxy and affixed with a 1 k$\Omega$ chip heater on the free end.
A matched pair of RuO bare-chip sensors, calibrated in separate experiments and mounted on thin Cu plates, were attached
to the specimen through 5 mil. diameter Au-wire thermal links bonded to the Cu plates and specimen with silver epoxy.
Measurements were performed in a $^3$He ``dipper'' probe with integrated superconducting solenoid. Magnetization measurements
were performed in separate experiments using a Quantum Design PPMS system. The dimensions, orientations, and demagnetization factors of all specimens can be found in the Supplemental Material \cite{SM}.
%
\begin{figure}[t]
\includegraphics[width=3.3in,clip]{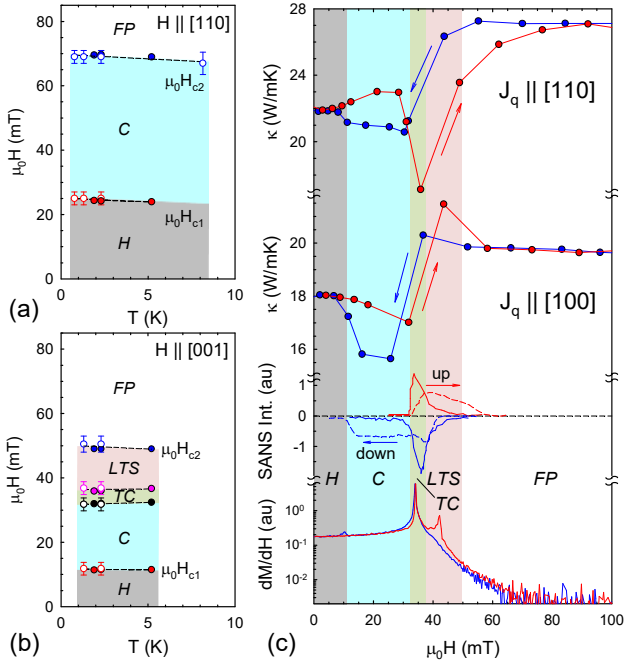}
\caption{Low-$T$ phase diagrams for (a) $H || [110]$ and (b) $H || [001]$. Filled (open) points are from features in $dM/dH$ ($\kappa$).
Dashed lines are guides. (c) Upper panels: the $H>0$ portion of $\kappa(H)$ scans from Fig.~\ref{Fig1}~(c) and (d). Middle panel:
SANS intensities for increasing (``up'') and decreasing (``down'') fields (from Ref.~\onlinecite{HalderPRB}) for the \emph{TC} (solid curves) and
\emph{LTS} (dashed curves) phases. Field values for the SANS data were shifted upward by 5~mT to align the onset of the
\emph{TC}-phase intensity with the leading edge of the sharp peak in $dM/dH$, consistent with the calibration established in
Ref.~\onlinecite{HalderPRB}. Lower panel: $dM/dH$ from Fig.~\ref{Fig1}~(d).}
\label{Fig2}
\end{figure}

Figure \ref{Fig1}~(a)-(d) show the field dependent (normalized) thermal conductivities, $\kappa(H)/\kappa(0)$, and average specimen
temperatures, $T_{avg}$ (upper and middle panels, respectively) for four different configurations of the applied field ($H$) and heat
flow ($J_q$): $H\perp J_q$ for (a), (c) and $H|| J_q$ for (b), (d). The derivative of the magnetization, $dM/dH$ (lower panels), was
determined from separate measurements on the same specimens (all field values represent internal fields, corrected for
demagnetization \cite{Demag}). The background shading distinguishes the spin phases. Three spin phases, helical (\emph{H}),
conical (\emph{C}), and fully polarized (\emph{FP}), are clearly reflected in the three measured quantities for
$H||\langle 110\rangle$ [Fig.~\ref{Fig1}~(a), (b)]. Note that the average specimen temperature is stable to within 3-4 mK
during measurements (conducted at a series of stabilized field values). Abrupt changes in $T_{avg}$ at the spin-phase transitions
mirror those in $\kappa$ and reflect energy exchange between the spin system and lattice under the near-adiabatic conditions of the
measurements \cite{PrasaiK}.

Additional sharp features and strong hysteresis distinguish the field-dependent quantities for $H||\langle 100\rangle$ [Fig.~\ref{Fig1}~(c), (d)]
from those with $H||\langle 110\rangle$. These additional features are attributed to the tilted conical (\emph{TC}) and low-temperature
skyrmion (\emph{LTS}) phases, identified in SANS \cite{Chacon,Qian} as unique to the $H||\langle 100\rangle$ orientation.
Phase diagrams are plotted in Fig.~\ref{Fig2} based on features observed in the field dependencies of $\kappa$ and $dM/dH$; for the latter we
followed the calibration of $dM/dH$ against SANS established by Halder {\it et al.} \cite{HalderPRB}, discussed in more detail below.
Changes in $\kappa(H)$ upon traversing the various phase boundaries are the principal results and focus of this work.

For $H||\langle 110\rangle$ [Fig.'s~\ref{Fig1}~(a), (b)], we see that $\kappa$ increases sharply with increasing field upon crossing
the helical-conical phase boundary. This clearly reflects the elimination of domain boundary scattering of magnons as the multidomain
helical phase, with three mutually perpendicular spiral spin domains aligned along the $\langle 100\rangle$ directions, is converted
to a single domain with conical spiral axis along the applied field. Note that this step-like enhancement of $\kappa$ occurs in
Fig.'s~\ref{Fig1}~(a) and (b) regardless of the conical spiral direction transverse to or along the heat flow. Within the conical
phase $\kappa$ increases as the conical angle decreases, and for $H||[1\overline{1}0]$ abruptly {\it decreases} upon entering the
fully polarized state [Fig.~\ref{Fig1}~(a)]. A similar behavior was observed for the same field orientation with [111] heat flow
\cite{PrasaiK}. Though a larger intrinsic spin gap in the fully polarized state might explain these features, the specific heat
shows only a modest, $\lesssim 5$~\% decline through the \emph{C-FP} transition \cite{HalderPRB}, suggesting a more substantial
decline in the magnon mean-free path.

The more complex and hysteretic behaviors for $\kappa(H)$ with $H||\langle 100\rangle$ are correlated in Fig.~\ref{Fig2}~(c) against the
SANS \cite{HalderPRB} and magnetization data through a more detailed comparison of the positive field portions of Fig.'s~\ref{Fig1}~(c) and
(d); the SANS intensities for the \emph{TC} and \emph{LTS} phases for both ascending (``up'') and descending (``down'') field are reproduced
in the middle panel. The principle features of $\kappa(H)$ are: (1) in ascending field $\kappa$ increases gradually by $\sim 1$~W/mK upon
entering the conical phase for $J_q||[110]$, but decreases by the same amount for $J_q ||[100]$, (2) $\kappa$ sharply decreases in
ascending field upon entering the tilted conical phase for $J_q||[110]$, but increases sharply for $J_q ||[100]$, (3) for both
orientations $\kappa$ is maximal or approaches its maximum value in the \emph{LTS} phase, (4) in descending field for both orientations,
the conical phase $\kappa$ is lower than in ascending field and exhibits a step-like increase toward the zero-field value upon crossing
the \emph{C-H} phase boundary.
\begin{figure}[t]
\includegraphics[width=3.3in,clip]{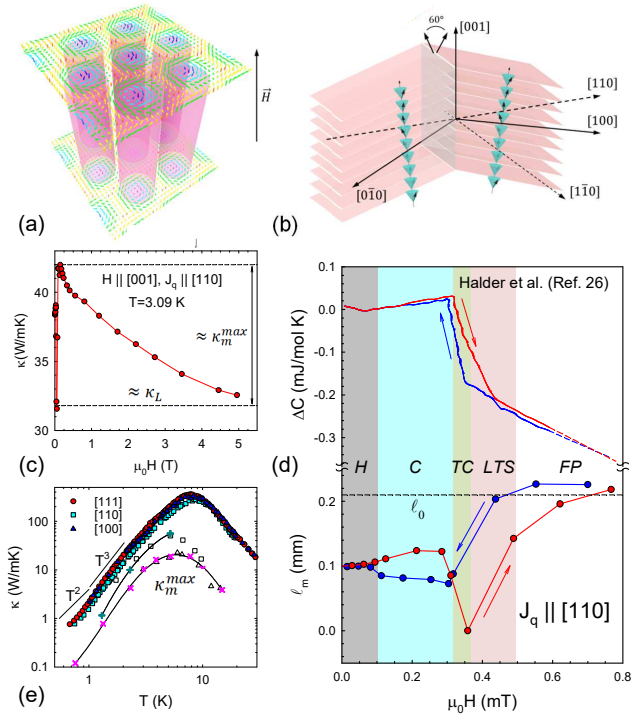}
\caption{(a) skyrmion spin textures in the \emph{LTS} phase, (b) an idealized planar boundary between tilted conical
domains of the \emph{TC} phase (the planes represent layers of atomic spins oriented in the direction dictated by the adjacent conical
spirals), (c) $\kappa (H)$ at $T=3.09$~K for a second specimen with
the same configuration as in Fig.~\ref{Fig1}~(c), but at higher field in the fully polarized state where $\kappa\to\kappa_L$ due to
the field-induced spin gap. Low-field data at additional temperatures for this specimen are presented in Fig.~S3 [\onlinecite{SM}],
(d) Change with field in the specific heat (from Ref.~\onlinecite{HalderPRB}) extended to higher field (dashed lines) and
magnon mean-free path computed from kinetic theory as described in the text from data in Fig.~\ref{Fig2}~(c). The horizontal
dashed line indicates the specimen transverse dimension $\ell_0$. (e) zero-field $\kappa(T)$ for heat flow along
different crystallographic directions. Also shown ($+,\ \times$ and solid curves as guides) are the maximum magnon contributions,
$\kappa_m^{max}$, inferred for two [110] specimens as described in the text, and for two [111] specimens from Ref.~\onlinecite{PrasaiK}
(open squares and triangles).}
\label{Fig3}
\end{figure}

Hysteresis, in all of the measured quantities in Fig.~\ref{Fig2}~(c), has its origin in the strongly first-order character of
the transition to the \emph{LTS} state which involves a large energy barrier requiring the creation
of pairs of Bloch points (magnetic monopoles) and is topologically protected \cite{Milde,Wild}. The \emph{LTS} phase is preceded in both
ascending and descending field by the \emph{TC} phase, but the \emph{LTS} phase persists to higher and lower fields, respectively. As a
result, the \emph{LTS} phase coexists with the conical phase in descending field and disappears only at the \emph{C-H} transition
(evidenced by the small peak in $dM/dH$). Clearly the long-range ordered skyrmion spin textures [Fig.~\ref{Fig3}~(a)] present minimal
scattering for magnons [feature (3) above] and thus suppressed values for $\kappa$ in the conical phase for descending field
[feature (4)] are attributed to spin disorder
associated with phase coexistence. Feature (1) might be related to channeling of spin waves with momenta transverse to the conical
spiral axis \cite{Janoschek,GarstRev}, but this possibility requires further investigation.

We focus the remainder of our discussion on the sharp drop in $\kappa$ upon entering the tilted conical phase for $J_q ||[110]$ [feature (2)],
interpreted to indicate that boundaries between tilted conical domains are extremely effective scatterers of magnons. The domain structure
within the \emph{TC} phase has not been established, but since the conical spiral axes tilt away from the $[001]$ field direction toward
$\langle 111\rangle$ directions (by angle $\phi\approx 30^{\circ}$ at $T=2$~K \cite{Chacon,Qian}), a planar boundary transverse to the
heat flow, as depicted in Fig.~\ref{Fig3}~(b) for two such domains, is possible. Even for a boundary with a more realistic transition
region of finite extent, the cross-sectional area of the domain boundary for $H||J_q||[100]$ would be substantially smaller, offering
an explanation for the absence of a downturn in $\kappa$ for this orientation. As noted in Ref.~\onlinecite{Qian}, the very small free
energy difference for conical spiral alignment along the $\langle 100\rangle$ and $\langle 111\rangle$ directions could lead to the
formation of a superstructure with a smooth rotation between tilted conical domains. Our results clearly favor a domain structure that
results in highly anisotropic magnon scattering, with strong scattering for magnon momentum \emph{transverse} to the applied field.

Data at higher field (up to 5~T) for a second $[110]$ specimen [Fig.~\ref{Fig3}~(c)] demonstrate that $\kappa_m$ is entirely suppressed
in the tilted conical phase for this orientation, since $\kappa(5~{\rm T})\to \kappa_L$ is comparable to the value of $\kappa$ at the
sharp drop upon entering the \emph{TC} phase. Further corroborating this interpretation, the computed magnon mean-free path $\ell_m$
for $J_q||[110]$ using data of Fig.~\ref{Fig2}~(c) achieves a maximum value [Fig.~\ref{Fig3}~(d)] in good
agreement with the effective transverse dimension of the sample. Also shown in Fig.~\ref{Fig3}~(d) are changes in the specific heat
with field $\Delta C(H)$ reported in Ref.~\onlinecite{HalderPRB}. We used kinetic theory, $\ell_m=3\kappa_m/(C_mv_m)$, where $\kappa_m$
was computed from the data assuming $\kappa=\kappa_L$ at the minimum, $v_m$ is the magnon velocity, and $C_m(H)$ was computed from
$\Delta C(H)$ (converted to volume units using 1~mol=$5.35\times 10^{-5} {\rm m}^3$) and assuming a magnetic specific heat per volume
from, $C_m(0)=(0.113/4)k_B(k_BT/D)^{3/2}$ with
spin-wave stiffness $D=52.6$~meV\ \AA$^2$ \cite{Portnichenko}. The dominant magnons for boundary-limited $\kappa_m$ have
$q_{dom}=(2.58k_BT/D)^{1/2}$ such that $v_m\simeq 1040 T^{1/2}$~m/s \cite{PrasaiK}.

Curves for $\kappa_m^{max}(T)$, determined from the difference between the minimum value and the maximum in the \emph{FP} state,
are shown in Fig.~\ref{Fig3}~(e) ($+$, $\times$ and solid curves) for both [110] specimens \cite{SM}; their magnitudes and temperature
dependencies, with peaks at $T\approx 5-6$~K, are consistent with prior results \cite{PrasaiK} on $\kappa_m^{max}(T)$ for $[111]$-oriented
specimens with $H||[1\overline{1}0]$ (open squares and triangles). The rapid decline in $\kappa_m^{max}$ at $T\gtrsim 6$~K for both
the [110] and [111] oriented specimens is the signature of other strong, intrinsic magnon scattering (e.g. magnon-magnon Umklapp or
magnon-phonon).

That a magnon heat conductivity as large as $\kappa_m^{max}\approx 60$~W/mK at 5~K [Fig.~\ref{Fig3}~(e) and Fig.~S2~(b) \cite{SM}] can
be suppressed to zero with a small change in applied field constitutes a novel, insulating spin-current switch \cite{SpinSwitch} that
could be exploited to control voltage readout in an interfacial heavy-metal thin film via the inverse spin-Hall effect. Since the
tilted conical phase arises from a competition between magnetic anisotropies generic to chiral magnets \cite{Chacon,Qian}, similar
phenomena might be found in other materials and temperature regimes.


In summary, the unprecedentedly large magnon thermal conductivity in Cu$_2$OSeO$_3$ provides a sensitive probe of the transitions
in applied field between its spin phases in the ground state. Our survey of various orientations of the heat flow and applied field
provide new insight into magnon transport within non-collinear spin phases, the influence of domain interfaces and phase coexistence
on magnon scattering,
and reveal a new mechanism by which spin currents in insulators may be manipulated.

\vskip .1in

This material is based upon work supported by the U.S. Department of Energy, Office of
Science, Office of Basic Energy Sciences, under Award Number DE-SC0008607
(UM). Work at JHU was supported as part of the Institute for Quantum
Matter, an Energy Frontier Research Center funded by the Department of Energy, Office of
Science, Office of Basic Energy Sciences, under Award Number DE-SC0019331, and
the NSF, Division of Materials Research, Solid State Chemistry, CAREER grant No.
DMR-1253562.

\clearpage
\onecolumngrid
\renewcommand*{\thefigure}{S\arabic{figure}}
\renewcommand*{\theequation}{S\arabic{equation}}
\renewcommand*{\thetable}{S\arabic{table}}
\setcounter{figure}{0}

\center{\Large\bf Supplementary Material for: \break Spin phases of the helimagnetic insulator Cu$_2$OSeO$_3$ probed by magnon heat conduction}

\section*{Specimen Details}

For all of the crystals described in this work Table I shows their
dimensions, effective cross-sectional areas, orientations of heat
flow and applied field during measurements, demagnetization
factors, and Figure labels in which their data appears.

%
\begin{table}[h]
\caption{Specimen details.}
\begin{ruledtabular}
\begin{tabular}{cccccc}
Dimensions (mm$^3$) & $\ell_0$~(mm)  & Heat Flow Dir. & Field Dir. & $N$ & Figure(s) \\
\hline
$2.08\times 0.26\times 0.17$ & 0.24 & $[110]$ & $[1\overline{1}0]$ & $0.39$ & 1~(a) \\
\hline
$2.06\times 0.26\times 0.13$ & 0.21 & $[110]$ & $[110]$ & $0.04$ & 1~(b), S1 \\
'' & '' & $[110]$ & $[001]$ & $0.33$ & 1~(c), 2~(c), S2~(a), (b) \\
\hline
$1.43\times 0.11\times 0.09$ & 0.11 & $[100]$ & $[100]$ & $0.03$ & 1~(d), 2~(c), S2~(c) \\
\hline
$1.50\times 0.23\times 0.18$ & 0.23 & $[110]$ & $[001]$ & $0.52$ & 3~(c), S3 \\
\end{tabular}
\end{ruledtabular}
\end{table}

\section*{Additional low-$T$ $\kappa(H)$ data}

Figure~S1-S3 show $\kappa(H)$ data at additional temperatures for
specimens discussed in the main text.
\begin{figure}[b]
\includegraphics[width=6.7in,clip]{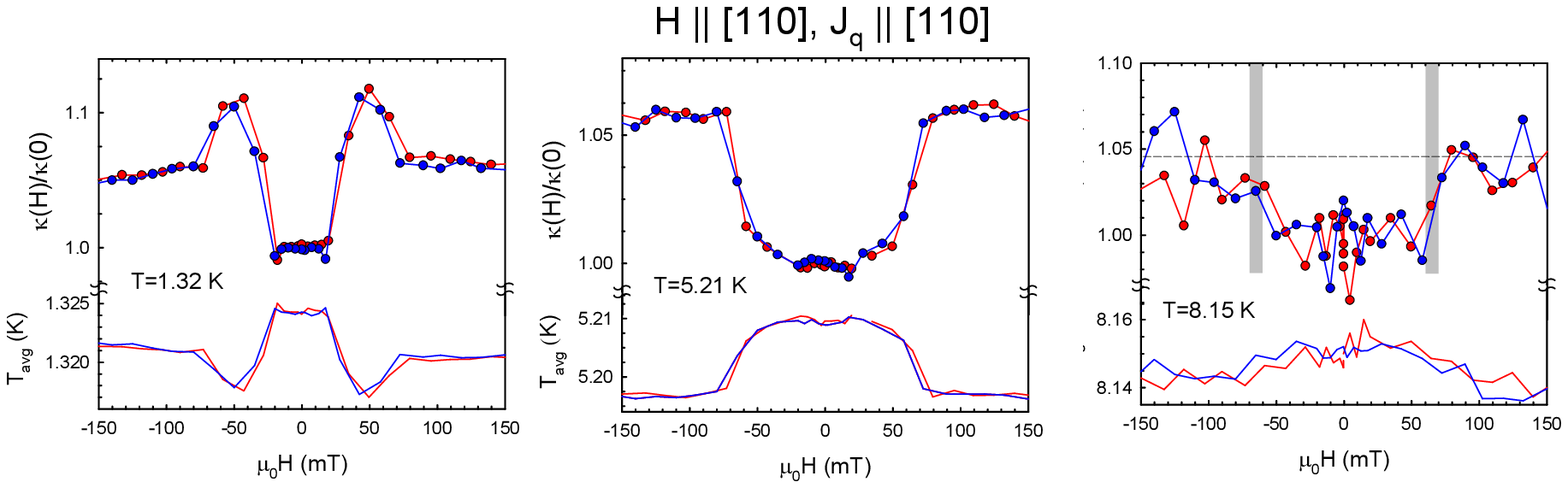}
\caption{$\kappa(H)/\kappa(0)$ at additional temperatures for the
$[110]$-oriented crystal from Fig.~1~(b). Shading in the last
panel reflects the uncertainty in determining $H_{c2}$ and the
error bar in the phase diagram [Fig.~2~(a)].} \label{S1}
\end{figure}
\begin{figure}[t]
\includegraphics[width=6.9in,clip]{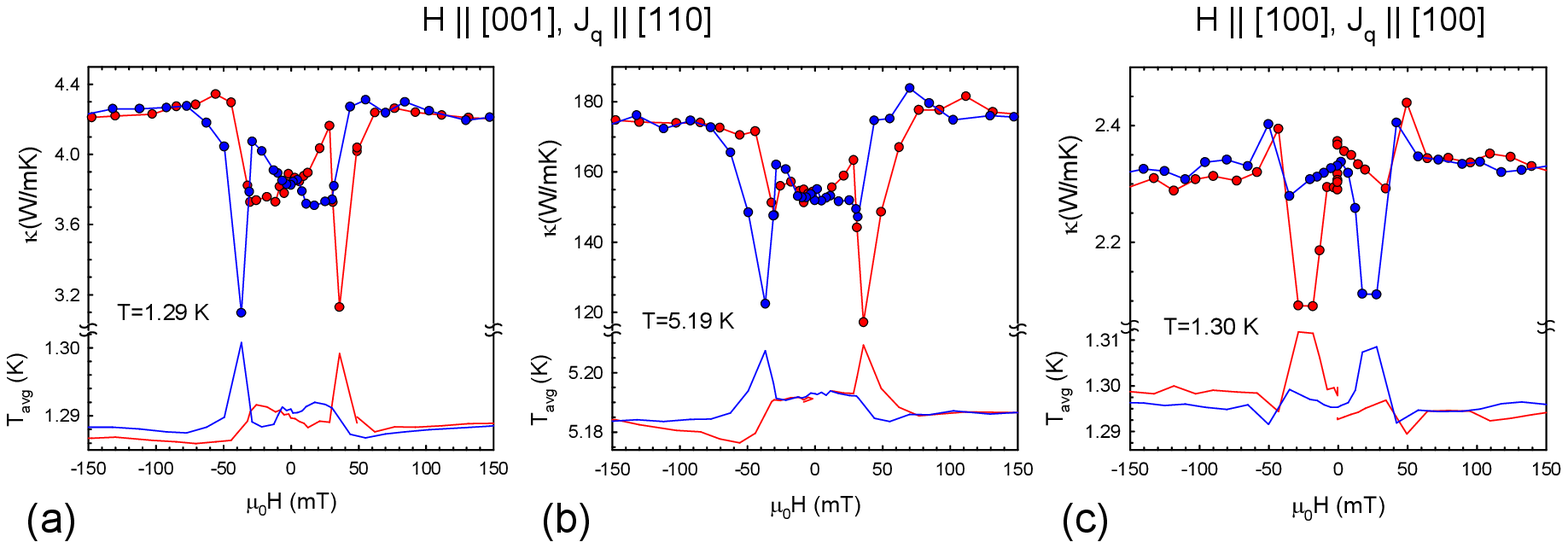}
\caption{$\kappa(H)$ at additional temperatures for crystals [(a)
and (b)] from Fig.~1~(c), and (c) from Fig.~1(d).} \label{S1}
\end{figure}
\begin{figure}[b]
\includegraphics[width=3.3in,clip]{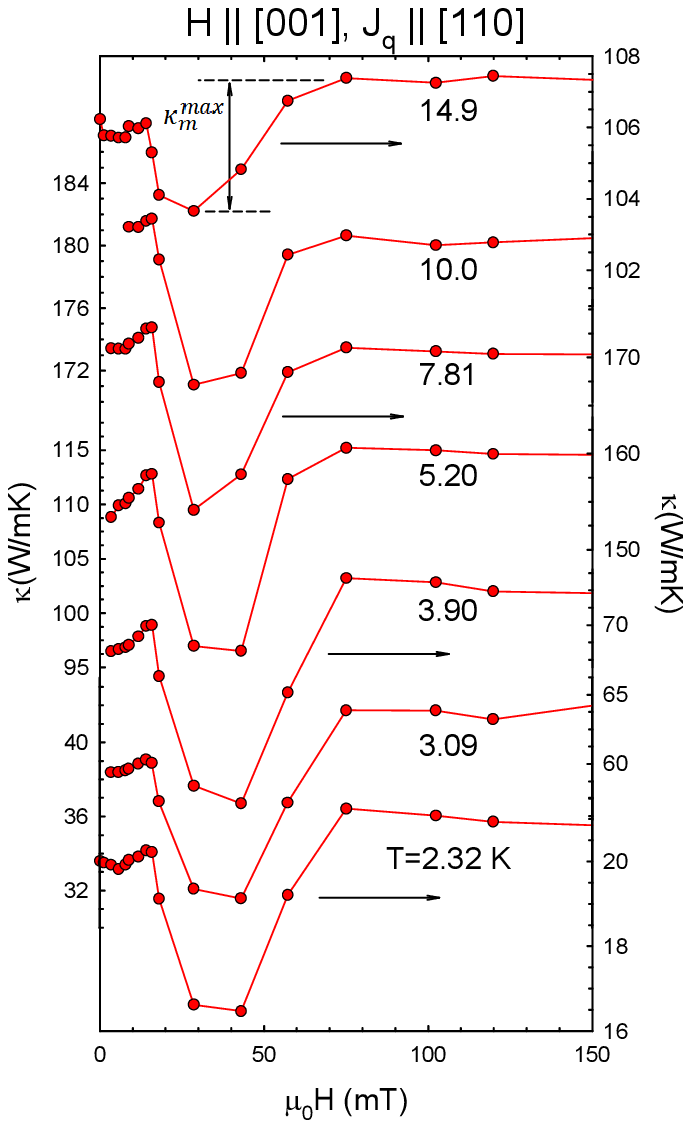}
\caption{$\kappa(H)$ at additional temperatures for the
$[110]$-oriented crystal from Fig.~3~(c). This specimen had a
maximum value $\kappa(T=8.5~{\rm K})$ about 30\% smaller than that
of the specimen from Fig.~(1)~(c), indicating a higher defect
density. This may contribute to the wider field region for the
``dip'' evident in $\kappa(H)$ (e.g. due to domain pinning), along
with field inhomogeneity due its larger demagnetization factor.}
\label{S2}
\end{figure}

\end{document}